%% file: version2012.tex
\documentclass[11pt]{article}
\usepackage{epsf}
\usepackage{amssymb}
\usepackage{amsfonts}
\usepackage{amsmath}
\title{May vortices produce a mass gap in $2D$ spin models \\
at weak coupling \thanks{Work supported
by Grant of Austrian Ministry of Science, Research
and Arts BM:WFK GZ 45.403/3-IV/3a/}}

\author{Oleg Borisenko\footnote{email: oleg@ap3.bitp.kiev.ua} and
Peter Skala\footnote{email: skala@kph.tuwien.ac.at}\\
Institut f\"ur Kernphysik, Technische Universit\"at Wien\\
A--1040 Vienna, Austria}

\begin{document}

\maketitle

\begin{abstract}
We consider the $2D$ $SU(N)$ principal chiral model and discuss a vortex
condensation mechanism which could explain the existence of a non-zero
mass gap at arbitrarily small values of the coupling constant. The
mechanism is an analogue of the vortex condensation mechanism of
confinement in $4D$ non-Abelian gauge theories. We formulate a
sufficient condition for the mass gap to be
non-vanishing in terms of the behaviour of the vortex free energy.
The $SU(2)$ model is studied in detail. In one dimension we calculate 
the vortex free energy exactly. An effective model for the center 
variables of the spin configurations of the $2D$ $SU(2)$ model
is proposed and the $Z(2)$ correlation function is derived 
in this model. We define a $Z(2)$ mass gap
in both the full and effective model and argue that they
should coincide whenever the genuine mass gap is non-zero.
We show via Monte-Carlo simulations of the $SU(2)$ model 
that the $Z(2)$ mass gap reproduces the full mass gap with 
perfect accuracy. We also test this mechanism in the positive link 
model which is an analogue of the positive plaquette model
in gauge theories and find excellent agreement between 
the full and the $Z(2)$ mass gap.
\end{abstract}

\newpage

\section{Introduction: Motivation and Problems}

The nature of a non-zero string tension (ST) at any value of the bare coupling
constant is one of the most striking puzzles of QCD. An analogous question 
exists in two-dimensional ($2D$) non-Abelian spin models with continuous
symmetry group where the mass gap (MG) is expected to be non-zero at
any temperature. Despite huge efforts in both cases, a solution of
this problem has not yet been found. 
A promising conjecture which could explain the existence of a non-zero
ST in QCD and thus permanent confinement of quarks is based on a
vortex condensation mechanism. In this article we adjust the lattice
formulation of this mechanism \cite{mack}
to the case of $2D$ non-Abelian spin models. 
Though this paper is in a spirit of \cite{kovacs},
our treatment of the problem differs in some aspects. We propose
a simple effective model for the $Z(2)$ degrees of freedom in the $SU(2)$ 
principal chiral model which explains how this mechanism could work 
in the weak coupling region of the lattice model.

The idea that condensation of vortices may be responsible for
confinement of static quarks in non-Abelian gauge theories with 
non-trivial center appeared already in the late
seventies \cite{hooft,yon}. The essential concept was
taken from $Z(N)$ lattice gauge theory (LGT) where the Wilson loop 
was known to obey an area law in the strong coupling region.
Since $Z(N)$ forms the center of the $SU(N)$ group
it was suggested that $Z(N)$ vortices can be also present in the more 
complicated $SU(N)$ theory and play an essential role in generating a
non-zero ST. However, $Z(N)$ LGT undergoes a phase transition
at weak coupling to the deconfinement phase with the Wilson loop
obeying a perimeter law. Since $Z(N)$ vortices in $Z(N)$ 
LGT may have a thickness of only one lattice spacing 
its contribution to the free energy becomes negligible 
in the weak coupling region where the system is well ordered at small
distances. What concerns $SU(N)$ LGT, 
attempts to calculate an effective $Z(N)$ theory at small coupling
making a perturbative expansion around $Z(N)$ solutions of the Yang-Mills
equations only lowered the critical coupling but did not remove 
it to zero as is expected to be the case for the correct confinement mechanism
\cite{yon}. 

In theories with continuous symmetry group like $SU(N)$, vortices
may have, however, a thickness of not only one but many lattice spacings.
Hence, there is the possibility to generalize the naive mechanism
of $Z(N)$ models and include all the possible vortex configurations
present in $SU(N)$ theories. Such a theory of confinement was
developed in \cite{mack} where also a theorem was proved which
makes a link between the behaviour of the Wilson loop and the vortex 
condensate. In this theory the definition of a vortex is 
actually not important. The only important issue is a change of
vorticity, and the vortex condensate is defined as the free energy
of such a change introduced by special singular $Z(N)$ gauge 
transformations. Over large distances typical configurations 
look like $Z(N)$ vortices, i.e. the basic field variables separated by 
such a distance are rotated by a $Z(N)$ element relatively to each other. 
The vortex condensation mechanism is
manifestly gauge invariant and presumably gives a nice explanation 
of the coexistence of confinement at large scales and perturbative 
behaviour at short range. 
The mechanism was investigated in many papers and some results
supporting its validity were found \cite{munst}-\cite{tomb}:

\begin{itemize}

\item
A strong coupling expansion of the vortex free energy up to the 12-th
order demonstrates that vortex configurations produce a
ST which coincides with the full ST
up to this order \cite{munst}. This result is gauge independent.

\item
The $Z(N)$ Wilson loop was shown to carry all the ST
to all orders of the strong coupling expansion \cite{gopf},
at least in the electric gauge. It was argued that $Z(N)$ Wilson
loops in different gauges differ only by perimeter contributions.

\item
If the magnetic-flux free energy vanishes in the limit of a large
uniform dilatation of a torus, the vortex free energy always
decreases exponentially. It is sufficient to produce confinement.
Using this property one can rigorously prove the lower confining
bound in three-dimensional $U(N)$ LGT \cite{yon1}.

\item
Simple intuitive ideas as well as some analytical results show
that a vortex mechanism is likely to lead to confinement at weak couplings
also in the positive plaquette model and the Mack-Petkova model,
which eliminates certain $Z(N)$ magnetic monopoles 
in $SU(N)$ LGT \cite{piet}.

\item
The technique to evaluate the contribution of vortices
of arbitrary thickness to the expectation value of any observable 
was developed in \cite{mw}. Such a ``preaveraged'' Wilson loop, i.e. 
calculated solely on the vortex contributions, exhibits
confining behaviour. While not rigorous, this result potentially
refers also to the weak coupling limit of the $3D \ SU(2)$ model.

\item
It has been proved that even the classical, though not naive limit
of $SU(N)$ LGT includes bare vortices in the continuum Lagrangian.
They are labeled by the nontrivial center elements of $SU(N)$
and are supported on closed $2D$ surfaces 
in four dimensions \cite{stul}. A one-loop expansion in a particular
background of such vortices shows instability of the vacuum  
implying that vortices must condense (become ``fatter'') in the
quantum theory already at two loop order.

\item
Recently, the vortex condensation mechanism of confinement was studied 
in the so-called maximal center gauge of $SU(2)$ LGT \cite{z2max}. 
The authors of these papers claimed that the Wilson loop computed with 
the center projected $Z(2)$ gauge field degrees of freedom  
carries almost the whole asymptotic ST. 
Consequently, excluding all $Z(2)$ vortices identified after
projection leads to a vanishing ST. 

\item
The $SU(2)$ partition function can be rewritten 
in the form of coupled $SU(2)/Z(2)$ and $Z(2)$ models which allows
to give a proper interpretation of different $Z(2)$ excitations
in the original model \cite{tomb}. 
Using plausible assumptions one can establish a link
between these excitations and the behaviour of the sign of the trace of
the Wilson loop. It was shown via Monte-Carlo (MC) simulations
that the sign of the trace of the Wilson loop carries all the 
information about the asymptotic behaviour of the fundamental string
tension.

\end{itemize}

It should be stressed that
earlier papers on the vortex condensation mechanism of confinement 
in continuum QCD cannot be regarded as an explanation of confinement 
\cite{corn}: In absence of a non-perturbative definition the
introduction of vortex configurations into the QCD Lagrangian seems
to be a completely ad-hoc procedure. Thus such models are not able to
give an explanation in terms of dynamical reasons why vortices should
become fat, i.e. why they are condensed. 

In general, there are two ways of looking at the vortex condensation
mechanism in lattice QCD. The first one
results from the desire to find an analogy with the continuum theory
and interprets the vortices responsible for confinement as an analogue
of the Nielsen-Olesen vortices and the QCD vacuum as the so-called 
spaghetti vacuum.
The second approach is based on the similarities between lattice QCD
and $2D$ non-Abelian spin models
where, while not so close to the continuum, one can give precise 
mathematical definitions to all quantities involved. Following the
latter idea, Mack and Petkova \cite{mack} formulated a condition which
could be called confinement mechanism by a vortex condensate. 
We would like to emphasize that while the relation of this approach 
to the spaghetti vacuum is rather vague at the present stage of
affairs the connection to $2D$ spin models is straightforward since one has 
precise definitions in both cases. Moreover, a quantity like the
vortex condensate is expected to be a genuine non-perturbative
quantity and thus it should be clear that it can be given a precise
meaning only in a non-perturbative approach such as the formulation of
QCD on a lattice. 

In this article we consider the vortex mechanism in some details on
the example of $2D$ spin models. The reason to deal with these models is the
following: The nature of the MG is
unknown despite the claim in the literature that its exact physical 
value is known. A vortex condensation mechanism is one possible 
candidate to explain the phenomenon of a non-zero MG. Moreover,
this mechanism has an analogue in gauge theories.
In view of all similarities between $2D$ spin models and $4D$ gauge
theories we think it is useful and instructive to study this mechanism 
on the example of simpler $2D$ models.

This paper is organized as follows. 
As a first point we give a definition of the vortex free
energy in terms of the original spin configurations. 
We introduce a vortex container which has the topology of a ring
in $2D$ and which is specified by certain boundary conditions (BC).
We prove a sufficient condition for the correlation function to
decrease exponentially which is precisely the analogue of the Mack-Petkova
theorem in LGT. This is done is Section 2.

Having identified the exponential decay of the vortex free energy
as a sufficient condition for producing a non-vanishing
MG the question arises what are the 
configurations of the spin field responsible for this 
rapid change of vorticity. 
Following the ideas of \cite{gopf,mack1} one can introduce a 
$Z(N)$ correlation function which measures the effect of vortices
on the full correlation function. In spin models with global
symmetry this is conceptually easier since we do not have to 
fix a gauge to uniquely define such a quantity. One can give the
corresponding arguments \cite{mack1} that this $Z(N)$ correlation
function defines completely the large distance behaviour of the full
correlation if a vortex mechanism is responsible for the
non-vanishing MG. We shall construct such a correlation and
derive an effective Ising-like model for $Z(2)$ excitations in Section 3.
Furthermore, we calculate the correlation function numerically for
the case of 
the $SU(2)$ principal chiral model using MC simulations. 
We find that the MG extracted from the $Z(2)$ correlation
function agrees almost completely with the asymptotic MG of the $SU(2)$ model.
A more delicate question is what one can expect in the case of the positive
link (PL) model where all the thin vortices are  eliminated. 
We introduce this model in analogy to the PP model
\cite{piet} and study it in the $SU(2)$ case. In particular, we
define and calculate $Z(2)$ correlations in the same way as in the full 
$SU(2)$ model. All these questions are subject of Section 4. 

Our conclusions and discussion are presented in Section 5.

\section{Vortex mechanism in spin models}

\label{secvort}

In this section we give a precise formulation of the vortex
condensation mechanism in two-dimensional spin models. A first hint
that such a mechanism could be crucial for the existence of a non-zero
MG came from the famous paper by Dobrushin  
and Shlosman \cite{spin}. These authors pointed out that the 
Mermin-Wagner theorem on the absence of spontaneous magnetization 
in two-dimensional spin systems follows from the intuitive idea that
in such systems long Peierls contours cost only little free energy by
making them thick. This is possible because the spins can rotate
slowly due to the continuous nature of the symmetry group.

In the following we derive a sufficient condition for the MG in
two-dimensional $SU(N)$ spin models to be non-vanishing. Our
derivation follows closely the one of the corresponding condition in
LGT by Mack and Petkova \cite{mack}. Nevertheless, we choose to adduce
it here because of pedagogical reasons and to make this paper
self-contained. 

For definiteness we consider the $SU(N)$ principal chiral model
on a two-dimensional periodic lattice $\Lambda^D$. 
Its partition function is given by
\begin{equation}
\label{1}
Z(\beta) =  \int \prod_{x \in \Lambda^D} D \mu (U_x) 
\exp [ \, \beta \sum_l \mbox{ReTr} (U_x U_{x+n}^\dagger) ], \qquad U_x \in SU(N).
\end{equation}
$\beta$ is the coupling constant, $D \mu$ denotes the normalized Haar
measure on $SU(N)$ and the integration extends over all sites $x$ of
the lattice. The sum in the exponent is a sum over all
links $l \equiv (x,n)$ of $\Lambda^D$. According to the
conventional scenario, the fundamental correlation function of spins
separated by a distance $R$ 
\begin{equation}
\label{2}
\Gamma_R(\beta) = \langle \, \mbox{ReTr} (U_0 U_{R}^\dagger) \, \rangle
\end{equation}
is expected to decrease exponentially at
any value of the coupling constant $\beta$ if $R$ is sufficiently large
\begin{equation}
\label{3}
\Gamma_R(\beta) \sim \exp(-m_c(\beta)R),
\end{equation}
with $m_c(\beta)$ the MG.

\begin{figure}[t]
\centerline{\epsfxsize=16cm \epsfbox{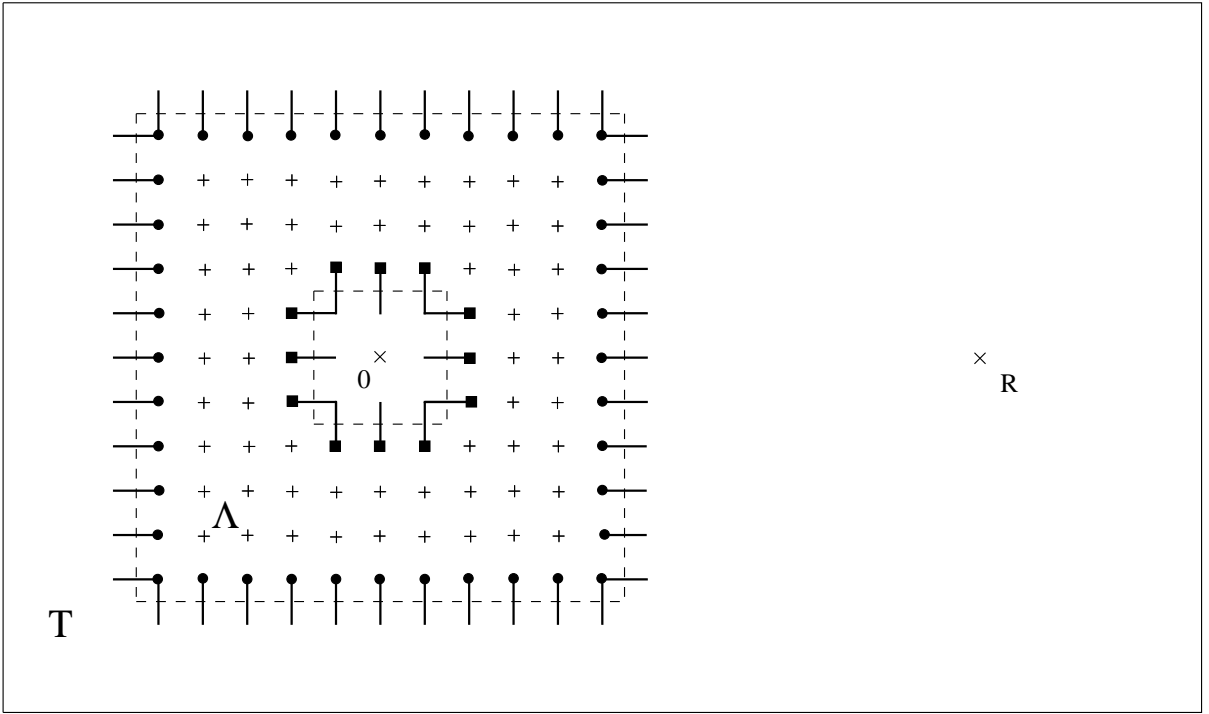}}
\hspace{1cm}
\caption{\label{cont}An example of a vortex container $T$ on a two-dimensional
lattice. The vortex container is the set of all lattice sites which
are enclosed by the two closed loops depicted as dashed lines,
i.e. the set of all crosses {\scriptsize $+$}, circles $\bullet$ and squares
{\tiny $\blacksquare$}. The interior $\Lambda$ of the container $T$ is the set 
of all crosses {\scriptsize $+$}, whereas the boundary $\partial T$ consists of all 
circles $\bullet$ (= the outer boundary $\partial T^{out}$) and squares {\tiny
$\blacksquare$} (= the inner boundary $\partial T^{in}$).}
\end{figure}
Let us consider now sets of links on $\Lambda^D$ which form closed 
loops on the dual lattice as shown in fig.~\ref{cont}. Following
\cite{mack} we call the region of the original lattice enclosed by
two such loops a vortex 
container $T$ which has the topology of a ring in two dimensions. We
arrange many such containers between the lattice sites $0$ and
$R$. Different containers may touch but not intersect each other. Let
$\partial T_i$ be the boundary (the set of
circles $\bullet$ and squares {\tiny $\blacksquare$} in
fig.~\ref{cont}) and $\Lambda_i$ the interior (the set of crosses
{\scriptsize $+$} in fig.~\ref{cont}) of the $i^{th}$ container
$T_i$. We define the complement $\Lambda_c$ of $\Lambda^D$ as 
$\Lambda_c = \Lambda^D / \prod_i \Lambda_i$. To make the following
derivation more transparent we rename the $SU(N)$ variables
$U_x$. Spins belonging to one of the vortex containers $T_i$ are renamed
$U'_x$, spins lying in the complement $\Lambda_c$ $\bar{U}_x$.

We start by rewriting the path integral expression for the fundamental
correlation function (\ref{2}) in terms of the variables $U'_x$ and $\bar{U}_x$
\begin{eqnarray}
\label{gamma}
\Gamma_R(\beta) \; &=& \frac{1}{Z} \int \prod_{x \in \Lambda_c}
D\mu(\bar{U}_x) \, \mbox{ReTr}  (\bar{U}_0 \bar{U}_{R}^\dagger) \, \exp
\left [ \, \beta \sum_{l \in \Lambda_c} 
\mbox{ReTr} (\bar{U}_x \bar{U}_{x+n}^\dagger) \right ]
\nonumber \\
&\cdot& \prod_i \left\{ \int \prod_{x \in T_i} D\mu(U'_x) \, \exp
[ \, \beta \sum_{l \in T_i}{}^\prime \, \mbox{ReTr} (U'_x U_{x+n}^{\prime\dagger}) ]
\prod_{x \in \partial T_i} \delta(U'_x \bar{U}_x^{-1}) \right\}.
\end{eqnarray}
The group $\delta$-function is necessary to avoid double integration
over spins defined on the boundary $\partial T_i$ of one of the
containers. The product $\prod_i$ runs over all
containers arranged between the lattice sites $0$ and $R$. In the sum
$\sum'$ in the exponent of the inner integral, links in the boundary
of the container $T_i$ are omitted. The inner integral equals the
partition function $Z(T_i,\bar{U}_x)$ defined on the container $T_i$ 
with boundary conditions $\bar{U}_x, \; x \in \partial T_i$,
\begin{equation}
\label{Zcont}
Z(T_i,\bar{U}_x) = \int \prod_{x \in T_i}
D\mu(U'_x) \exp \left [ \, \beta \sum_{l \in T_i}{}' \, 
\mbox{ReTr} (U'_x U_{x+n}^{\prime\dagger}) \right ]
\prod_{x \in \partial T_i} \delta(U'_x \bar{U}_x^{-1}).
\end{equation}
An important property of $Z(T_i,\bar{U}_x)$ is the invariance under
$SU(N)$ transformations of the boundary conditions. In particular,
$Z(T_i,\bar{U}_x)$ remains unchanged under the transformation
\begin{equation}
\label{trafo}
\bar{U}_x \; \to \; \omega_i^{-1} \bar{U}_x, \qquad x \in \partial T_i,
\end{equation}
with $\omega_i$ an element of the center $Z(N)$ of $SU(N)$. Let us
perform now the following variable substitution
\begin{alignat}{2}
\label{varsub}
\bar{U}_x \; &=  \; U_x \prod_{k=i}^{N_c} \omega_k   \qquad \qquad \mbox{if}
\qquad  &
\Lambda_{i-1} < \; &x < \Lambda_{i} \nonumber \\
             &=  \; U_x \prod_{k=1}^{N_c} \omega_k &
                   &x < \Lambda_1 \nonumber \\
             &=  \; U_x &
\Lambda_{N_c}   < \; &x.
\end{alignat}
The notation $\Lambda_{i-1} < x < \Lambda_{i}$ means that the lattice
site $x$ belongs to the region which is enclosed by the interior
$\Lambda_{i-1}$ of the container $T_{i-1}$ and the interior
$\Lambda_{i}$ of the container $T_{i}$. $N_c$ is the total number of
containers. In the special case of
$\bar{U}_0$ and $\bar{U}_R$, the substitution (\ref{varsub}) results in
$\bar{U}_0 = U_0 \prod_i \omega_i $ and $\bar{U}_R = U_R$. For the
correlation function (\ref{gamma}) we can write
\begin{equation}
\label{gamma2}
\Gamma_R(\beta) \; = \frac{1}{Z} \int \prod_{x \in \Lambda_c}
D\mu(U_x) \, \mbox{ReTr}  (U_0 U_R^\dagger) \, \exp
[ \, \beta \sum_{l \in \Lambda_c} \mbox{ReTr} (U_x U_{x+n}^\dagger) ]
\; \prod_i \omega_i Z(T_i,U_x^{\omega})
\end{equation}
with boundary conditions $U_x^{\omega}$ on the container $T_i$
\begin{alignat}{2}
\label{bcond1}
U_x^{\omega} \; &=  \; U_x \prod_{k=i+1}^{N_c} \omega_k   \qquad \qquad \mbox{if}
\qquad  & x &\in \partial T_i^{\text {in}} \nonumber \\
                &=  \; U_x \, \omega_i^{-1} \prod_{k=i+1}^{N_c} \omega_k & x
&\in \partial T_i^{\text {out}}. 
\end{alignat}
$\partial T_i^{\text {in}}$ and $\partial T_i^{\text {out}}$ are the
inner (the set of squares {\tiny $\blacksquare$} in fig.~\ref{cont})   
and outer boundary (the set of circles $\bullet$ in fig.~\ref{cont}) 
of the container $T_i$ respectively. We
further simplify the boundary conditions (\ref{bcond1}) by applying the
transformation $U_x \to U_x \, \omega_i \prod_{k=i+1} \omega_k^{-1}$
under which $Z(T_i,U_x^{\omega})$ remains unchanged. Thus we finally can
write 
\begin{alignat}{2}
\label{bcond2}
U_x^{\omega} \; &=  \; U_x \, \omega_i   \qquad \qquad \mbox{if}
\qquad  & x &\in \partial T_i^{\text {in}} \nonumber \\
                &=  \; U_x & x &\in \partial T_i^{\text {out}}
\end{alignat}
for the boundary conditions $U_x^{\omega}$ of the container
$T_i$. Since $\omega_i \in Z(N)$ is arbitrary, we may sum (integrate)
over $\omega_i$ in (\ref{gamma2}) using the normalized Haar measure on $Z(N)$.
Using the trivial relations
\begin{eqnarray}
\label{rela}
\left| \mbox{ReTr}(U_0 U_{R}^\dagger) \right| \; &\leq&  \; \mbox{Tr}({\bf 1 })
  \nonumber \\ 
\left| \frac{\sum_{\omega_i} \omega_i Z(T_i, U_x^{\omega}) }
{\sum_{\omega_i} Z(T_i, U_x^{\omega})} \right| \; &\leq&  \; \max_{U_x} \left| 
\frac{\sum_{\omega_i} \omega_i Z(T_i, U_x^{\omega}) }{\sum_{\omega_i} 
Z(T_i, U_x^{\omega})} \right|
\end{eqnarray}
and with the help of the identity
\begin{equation}
\label{identity}
Z \; = \; \int \prod_{x \in \Lambda_c}
D\mu(U_x) \exp [ \, \beta \sum_{l \in \Lambda_c} \mbox{ReTr} (U_x
U_{x+n}^\dagger) ] \; \prod_i \sum_{\omega_i} Z(T_i,U_x^{\omega})
\end{equation}
we derive the following bound for the correlation function
(\ref{gamma2})
\begin{equation}
\label{gamma3}
\left| \Gamma_R(\beta) \right| \; \leq \; \mbox{Tr}({\bf 1}) \prod_i \max_{U_x} 
\left| \frac{\sum_{\omega_i} \omega_i Z(T_i, U_x^{\omega}) }{\sum_{\omega_i} 
Z(T_i, U_x^{\omega})} \right|.
\end{equation}
The maximum on the right hand side of (\ref{gamma3}) is to be
understood as the maximum under the boundary conditions (\ref{bcond2}).
In order to make this result more transparent we will specify it for
the case of $SU(2)$. Then $\omega_i$ can take values $\pm1$ and the
bound (\ref{gamma3}) can be written
\begin{equation}
\label{CRineq}
\left| \Gamma_R(\beta) \right| \; \leq \; \mbox{Tr}({\bf 1}) \prod_i \max_{U_x}
\left| V_i(T_i,U_x^{\omega}) \right|,
\end{equation}
where we introduced
\begin{equation}
\label{VI}
V_i(T_i,U_x^{\omega}) \; = \; \frac{1 - q_i}{1 + q_i}.
\end{equation}
$q_i$ is called the vortex free energy and is given by the ratio
of partition functions defined on the vortex container $T_i$ with
boundary conditions $U_x^{\omega}$ (\ref{bcond2})
\begin{equation}
\label{Vfren}
q_i = \frac{Z(T_i,U_x^{\omega = -1})}{Z(T_i,U_x^{\omega = 1})}. 
\end{equation}
The physical meaning of $q_i$ (and more generally of $V_i$) is rather
obvious: $q_i$ measures the change of free energy $\exp(-\Delta F)$
introduced by a singular transformation $U_x \to U_x \omega_i$ on one
of the boundaries of the container $T_i$. Let $L_i$ be the diameter of
the container $T_i$. Then we say that the transformation $U_x \to U_x 
\omega_i$ performed on one of the boundaries of the container introduces a
vortex of thickness $L_i$ into the system and $q_i$ measures the
energy needed to create such a vortex.

We are now ready to explain what we call vortex condensation mechanism
for generating a non-zero MG in $2D$ $SU(N)$ spin models. Let
$V_i^{max}$ be the maximum of $| V_i |$ under the boundary conditions
$U_x^{\omega}$ defined in (\ref{bcond2}); i.e. $V_i^{max} = \max_{U_x}
| V_i |$. Suppose that there exists such a mass $m_v$ that for each
container $T_i$ and for sufficiently large diameter $L_i$, 
the maximum $V_i^{max}$ behaves according to
\begin{equation}
\label{fren1}
V_i^{max} \; \sim \; \exp(-m_v (\beta) L_i).
\end{equation}
Then, with $R = \sum_i L_i$, the bound (\ref{CRineq}) for the
correlation function reads
\begin{equation}
\label{ineq1}
\left| \Gamma_R(\beta) \right| \; \leq \; \mbox{const} \cdot \exp(-m_v (\beta) R), 
\end{equation}
which is the expected exponential decay. We term $m_v$ vortex MG
to distinguish it from the genuine MG $m_c$ extracted from the
correlation function. If $m_v = m_c$, condensation of vortices can be
made responsible for generating a non-zero MG in $2D$ $SU(N)$ spin
models.

\begin{figure}[t]
\centerline{\epsfxsize=8cm \epsfbox{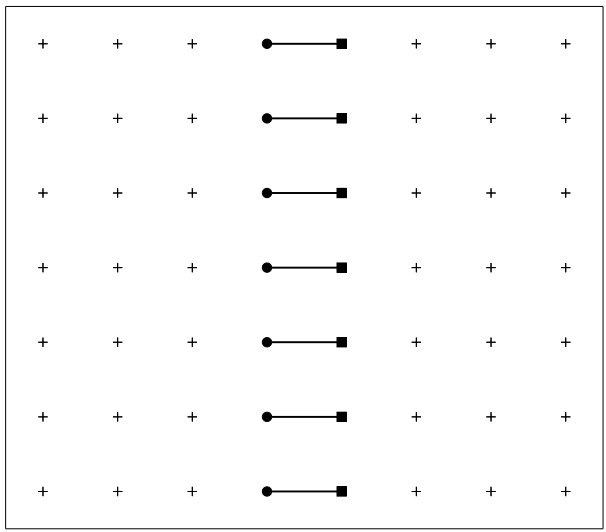}}
\hspace{1cm}
\caption{\label{cont2} A lattice (periodically closed) which has the 
same topology as the container shown in fig.~\ref{cont}. Spins defined
on circles $\bullet$ and squares {\tiny $\blacksquare$} are fixed to
given values. In the case of the vortex condensation mechanism
introduced in ref.~\cite{kovacs}, the coupling $\beta$ is changed to
$-\beta$ on the depicted links.}
\end{figure}
The crucial quantity in a vortex condensation mechanism is the vortex
free energy $q_i$ which is originally defined on the container $T_i$
but which may be calculated on any lattice having the same topology as
the container; i.e. the topology of a ring. Thus, we have to consider
a lattice with periodic boundary conditions in one direction and fixed
boundary conditions in the other direction. A corresponding lattice is
shown in fig.~\ref{cont2}. The vortex free energy $q$ on this lattice
is given by
\begin{equation}
\label{free2}
q = \frac{Z(U_x \, \omega)}{Z(U_x)},
\end{equation}
where in the denominator values $U_x$ are assigned to spins depicted as 
circles $\bullet$ and squares {\tiny $\blacksquare$} in
fig.~\ref{cont2}, while in the numerator spins living either on
circles or on squares are fixed to values $U_x \omega$, with 
$\omega \in Z(2)$. The partition function $Z$ is defined in (\ref{Zcont}).

Finally, we think that still some comments are needed at this place:
\begin{itemize}
\item
Configurations which rotate slowly from one center element to another
one on some characteristic length scale are termed ``thick
vortices''. According to the paper by Dobrushin and Shlosman \cite{spin}
they produce disordering effects which are sufficient to enforce the
long distance correlation function to fall off to zero at any
coupling; i.e. they guarantee the absence of magnetization. However,
a priori there is no reason to believe that thick vortices provide an
{\it exponential} fall off. It might be that for $\beta \to \infty$ these
configurations can only account for a power law decay of the
correlation function. Two conclusions can be drawn from such a
scenario. 1) Thick vortices are not responsible for a non-zero MG at
arbitrarily large values of $\beta$ or 2) they are at small $\beta$
but the system undergoes a phase transition to a massless phase at a
finite value of $\beta$. 
\item
The above vortex condensation theory does not give an exact definition
of a vortex. The only important issue is a change of vorticity over some
characteristic length scale. To specify a vortex completely, 
one has to define the rate of this change
and, possibly, the dependence of the characteristic length scale
on the bare coupling $\beta$. It is reasonable to assume that on this
length scale, the $SU(N)$ spin model looks like an effective $Z(N)$ model
for special $Z(N)$ excitations which one should be able to extract from the
original configurations.
\end{itemize}

\subsection{$1D$ model}

Before finishing this section we will demonstrate
on a simple example how the mechanism described above works in practice.
The simplest example we can imagine is the $SU(2)$ spin model
in one dimension where it is well known that the correlation function shows
an exponential fall off at any value of the coupling constant $\beta$
and the corresponding MG is given by
\begin{equation}
\label{1DMG}
m_c(\beta) = \ln \left( \frac{I_1(\beta)}{I_2(\beta)} \right),
\end{equation}
with $I_n$ modified Bessel functions. In one dimension
a vortex container is simply a chain of spins
with Dirichlet-like boundary conditions. More precisely:
let $L$ be the length of the $1D$ chain and let us fix 
the spins on the boundary to some arbitrary value 
%
\begin{equation}
\label{1DBC}
U_{x=0} = W_1,\quad U_{x=L} = W_2 .
\end{equation}
Define $W = W_1 W_2^\dagger$. On a finite lattice, the partition
function $Z$ can be calculated exactly
\begin{equation}
\label{1DZ}
Z(W) = \beta^{-L}\sum_{n=0,1/2,...}^{\infty} (2n+1) \chi_n(W)
\left[ I_{2n+1}(\beta) \right]^L ,
\end{equation}
where $\chi_n(W)$ is the character of the $n^{th}$ representation of $SU(2)$.
Perform now a nontrivial $Z(2)$ transformation on one of the boundary spins
of our chain, e.g. $W_2 \to \omega W_2 = - W_2$. The character
$\chi_n(W)$ transforms according to
\begin{equation}
\label{1Dchi}
\chi_n(W) \to \chi_n(-W) = (-1)^{2n}\chi_n(W).
\end{equation}
For the vortex free energy $q$ (\ref{Vfren}) we obtain
\begin{equation}
\label{q1D}
q = \frac{Z(-W)}{Z(W)} \approx 1 - 4 \chi_{1/2}(W)
\left( \frac{I_2(\beta)}{I_1(\beta)} \right)^L + ...,
\end{equation}
where the terms which vanish faster
in the thermodynamic limit have been neglected. 
The maximum of $V$ (\ref{CRineq}) occurs
at $W = {\bf 1}$ and we finally find
\begin{equation}
\label{V1D}
\max_W V \; \sim  \; \exp [-L \ln \left( \frac{I_1(\beta)}{I_2(\beta)}
\right) ].
\end{equation}
Thus, in the $1D$ model the change of $V$ introduced by the
creation of a thick vortex shows an exponential decrease with a MG
 which equals the MG extracted from the correlation function 
(\ref{1DMG}). 
Of course, in one dimension this result is in a sense trivial but it
shows nevertheless transparently how the idea described above works.

\section{Effective model for $Z(2)$ excitations}

In this section we address the question which are the spin field
configurations playing a crucial role in a vortex condensation theory.
In the case of gauge theories, it was suggested long ago that certain
$Z(N)$ excitations of the gauge field are responsible for an area law 
behaviour of the Wilson loop \cite{mack1}. Here, we present an
effective model for $Z(2)$ excitations in the $2D \ SU(2)$ spin model
and establish a link to the vortex condensation mechanism. The main 
assumption of this model is supported by MC data which will be 
presented in next Section.

As a first step we fix proper boundary conditions of the
two-dimensional lattice. It is convenient to take a
periodic lattice in $y$-direction and to fix the $SU(2)$ spins to
elements of the $Z(2)$ subgroup if $x=0$ and $x=L$. $L$ is the linear
extension of the lattice. It should be
stressed that fixing boundary conditions is not necessary for systems 
with global symmetry since the Mermin-Wagner theorem guarantees independence
of the results on BC in the infinite volume limit. In general, a dependence
on BC should vanish faster than any long-distance correlation function.
In our case fixing boundary conditions is only a question of
convenience and proper definitions.
%

We start by rewriting the partition function of the $SU(N=2)$
principal chiral model (\ref{1}) using the representation
\begin{equation}
\label{5}
U_x \, = \, z_x \bar{U}_x
\end{equation}
for the spin field $U_x \in SU(2)$, where $z_x$ is a $Z(2)$ element 
and $\bar{U}_x \in SU(2)/Z(2)$. In the new variables the BC for $z_x$ 
are free ones while for $\bar{U}_x$ we have to impose Dirichlet BC in
$x$-direction. For the invariant measure on the $SU(2)$ group we write
\begin{equation}
\label{6}
D \mu (U_x) = \frac{1}{2} \sum_{\{z_x\}=\pm 1} D \mu (\bar{U}_x),
\end{equation}
where $D \mu (\bar{U}_x)$ is an invariant measure on the $SO(3)$ group.
Let us recall that the invariant measure on the $SU(N)/Z(N)$ group
coincides with the $SU(N)$ measure up to the restriction
\begin{equation}
\label{o3meas}  
-\frac{2\pi}{N} \leq \arg(\mbox{Tr}U_x) \leq \frac{2\pi}{N}.
\end{equation}
In other words, in the invariant 
$\bar{U}_x$-integration the trace of
the fundamental characters is restricted to positive values. 
Performing now the summation over the $Z(2)$ elements $z_x$ one can rewrite
the partition function (\ref{1}) as (up to an irrelevant constant)
\begin{equation}
\label{Isingrep}
Z(\beta) =  \sum_{\{s_l\}=\pm 1} \sum_{\cal L} \int \prod_xD \mu (\bar{U}_x)
\exp [\, \beta \sum_{l} s_l \mbox{Tr} (\bar{U}_x \bar{U}_{x+n}^\dagger) ]
\prod_{l \in {\cal L}} s_l.
\end{equation}
${\cal L}$ is a set of closed loops and we have introduced a new
$Z(2)$ link variable $s_l$. This representation for $Z(\beta)$ (and a
similar one in the case of $SU(2)$ gauge theory) was derived and
investigated in \cite{noncm}. In terms of the link variable $s_l$  
the singular transformations discussed in section \ref{secvort} may be
defined as a change of the sign of $s_l$ on the boundaries of a vortex
container $T$ (only in the action).
The sum over ${\cal L}$ in (\ref{Isingrep}) is a sum over all closed loops 
(including all their possible products) taken with an appropriate weight. 
It is defined exactly in the same way as 
in the two-dimensional Ising model where much is known about the
properties of such a loop expansion. However, unlike the Ising model,
the coefficients defined on the links of closed loops are not constant.
Moreover, if we consider the model (\ref{Isingrep}) as an Ising-like model
with a fluctuating coupling constant, we observe that the coupling
is not positive definite (despite $\mbox{Tr}\bar{U}_x>0$ this is not the case
for the character of the product of two group elements).
At least at first glance, this could imply that fluctuations
of the link variable $s_l$ may persist down to weak
coupling and cause the disorder which is needed for an exponential
behaviour of the correlation function. In fact, this is precisely
what we are going to work out. 

As a first step we perform the sum over the link variables $s_l$. 
It results in
\begin{equation}
\label{Isingeff}
Z(\beta) =  \sum_{\cal L} \int \prod_xD \mu (\bar{U}_x)
\prod_l\cosh [\, \beta \, \mbox{Tr} (\bar{U}_x \bar{U}_{x+n}^\dagger) ]
\prod_{l \in {\cal L}}
\tanh [\, \beta \, \mbox{Tr} (\bar{U}_x \bar{U}_{x+n}^\dagger) ].
\end{equation}
For sufficiently large values of $\beta$ we obtain
\begin{equation}
\label{PFasym}
Z(\beta) =  \sum_{\cal L} (\tanh 2\beta)^{\mid {\cal L} \mid} 
\int \prod_xD \mu (\bar{U}_x)
\prod_l \exp [ \, \beta \, | \mbox{Tr} (\bar{U}_x
\bar{U}_{x+n}^\dagger)| \, ]
\prod_{l \in {\cal L}}\sigma_l  \  + {\cal O} ( e^{-4\beta} ),
\end{equation}
where
\begin{equation}
\label{sign}
\sigma_l = \mbox{sign} [ \mbox{Tr} (\bar{U}_x \bar{U}_{x+n}^\dagger) ].
\end{equation}
>From the last equations we see that at large values of $\beta$ the 
original partition function, and so the free energy, can be written as
a product of two partition functions - the partition function of a 
$SU(2)/Z(2)$ model and the partition function of an Ising-like model.
Precisely
\begin{equation}
\label{fctrz}
Z(\beta) = Z^{SU(2)/Z(2)}Z^I,
\end{equation}
with
\begin{equation}
\label{PFsz}
Z^{SU(2)/Z(2)} = \int \prod_xD \mu (\bar{U}_x) \prod_l
\exp [ \, \beta \, | \mbox{Tr} (\bar{U}_x \bar{U}_{x+n}^\dagger) | \, ]
\end{equation}
and the Ising-like partition function
\begin{equation}
\label{PFIsing}
Z^I =  \sum_{\cal L} (\tanh 2\beta)^{\mid {\cal L} \mid} F({\cal L}).
\end{equation}
$F(\cal L)$ is defined as an expectation value 
\begin{equation}
\label{FL}
F({\cal L}) = \langle \prod_{l \in {\cal L}} \sigma_l \rangle_{SU(2)/Z(2)}
\end{equation}
which has to be evaluated in the ensemble (\ref{PFsz}). In the case of
the fundamental correlation function (\ref{2}) for $N=2$ we follow the same
strategy and obtain for large $\beta$-values
\begin{eqnarray}
\label{CFsz}
\Gamma_R(\beta) = (Z^{SU(2)/Z(2)}Z^I)^{-1} 
\sum_{\cal P} (\tanh 2\beta)^{\mid {\cal P} \mid}
\sum_{{\cal L}/l\in {\cal P}} (\tanh 2\beta)^{\mid {\cal L} \mid}
\int\prod_xD\mu (\bar{U}_x) \, \mbox{Tr} (\bar{U}_0 \bar{U}_R^\dagger) \;
\times \nonumber \\
\exp [ \, \beta \, | \mbox{Tr} (\bar{U}_x \bar{U}_{x+n}^\dagger) | \, ] 
\prod_{l \in {\cal P}}\sigma_l \prod_{l \in {\cal L}}\sigma_l \
+ \  {\cal O}( e^{-4\beta} ).
\end{eqnarray}
Here, the sum over ${\cal P}$ is a sum over all paths (including
all their possible products) connecting the lattice sites $0$ and $R$. 
In addition, one has to sum up over all possible
closed loops on the lattice which must not have a link in common with a given
path ${\cal P}$. Again, this is in full accordance with the expansion of the
correlation function  
in the Ising model. To see this explicitly, let us write down the
corresponding formulae for the Ising model. For the partition function
one gets an expansion in terms of loops (up to a constant)
\begin{equation}
\label{Isingm}
Z^{{Ising}} =  \sum_{\cal L} (\tanh\beta)^{\mid {\cal L} \mid},
\end{equation}
while the correlation function reads
\begin{equation}
\label{Isingcor}
\Gamma_R^{{Ising}}(\beta) = (Z^{{Ising}})^{-1} \sum_P (\tanh\beta)^
{\mid {\cal P} \mid}
\sum_{{\cal L}/l\in {\cal P}} (\tanh\beta)^{\mid {\cal L} \mid}.
\end{equation}
These formulae have to be compared with the asymptotic expansions
(\ref{PFasym}) and (\ref{CFsz}) in the case of the $SU(2)$ model.
In the Ising model it is known that for $\beta$-values below
$\beta_c$ the correlation function is well
approximated by the expression
\begin{equation}
\label{Isingcorapp}
\Gamma_R^{{Ising}}(\beta < \beta_c) \approx 
\sum_{{\cal P}_c} (\tanh\beta)^{\mid {\cal P}_c \mid},
\end{equation}
where we have to take into account only the so-called 
connected graphs ${\cal P}_c$. In the following we shall show that one can
arrive at a similar expression for the correlation function in the
$SU(2)$ model by making suitable assumptions. Firstly, we introduce a
$Z(2)$ correlation function in the $SU(2)$ ensemble according to
\begin{equation}
\label{z2corr}
\Gamma_R^{Z(2)}(\beta) = \langle z_0 z_R \rangle  
\end{equation}
with $z_x = \mbox{sign} ( \mbox{Tr} U_x )$ and claim that it
reproduces the correct long distance behaviour of
the full $SU(2)$ model, i.e. the MG extracted from
(\ref{z2corr}) coincides with the full $SU(2)$ MG. This assumption
will be confirmed in the next section by numerical results. 
It leads to the following expression for the asymptotic expansion of
the $SU(2)$ correlation function (\ref{CFsz})
\begin{equation}
\label{su2corr}
\Gamma_R(\beta) = \frac{1}{Z^I} 
\sum_{{\cal P}} (\tanh 2\beta)^{\mid {\cal P} \mid}
\sum_{{\cal L}/l\in {\cal P}} (\tanh 2\beta)^{\mid {\cal L} \mid} \; 
\langle \prod_{l \in {\cal P}}\sigma_l 
\prod_{l \in {\cal L}}\sigma_l \rangle_{SU(2)/Z(2)},
\end{equation}
where the expectation value is defined in the ensemble (\ref{PFsz}).
Our next assumption is that for $\beta$-values smaller than $\beta_c$ 
(\ref{su2corr}) can be written as a sum over connected graphs like in
the Ising model
\begin{equation}
\label{CFszapp}
\Gamma_R(\beta < \beta_c) \approx  
\sum_{{\cal P}_c} (\tanh 2\beta)^{\mid {\cal P}_c \mid} \; 
\langle \prod_{l \in {\cal P}_c}\sigma_l \rangle_{SU(2)/Z(2)}.
\end{equation}
This formula has to be compared with the corresponding formula
(\ref{Isingcorapp}) in the case of the Ising model.
In the $SU(2)$ case the conventional scenario means $\beta_c=\infty$. Thus we  
expect that approximation (\ref{CFszapp}) works rather well in the whole
region of the bare coupling. Of course, its goodness is 
determined by the behaviour of the expectation value on the right hand
side of (\ref{CFszapp}).
Suppose for a while that $\langle \prod_{l \in {\cal P}_c}\sigma_l
\rangle_{SU(2)/Z(2)}$
is small enough to cancel the fast growing number
of connected paths with length $|{\cal P}_c|$ and to guarantee
a fast convergence on the right-hand side of (\ref{CFszapp}). 
Then, it is easy to give a prediction for the MG.
To leading order one finds for $\beta < \beta_c$
\begin{equation}
m_{eff}(\beta < \beta_c) \approx -\frac{1}{R} \ln \left[ \, (\tanh 2\beta)^R
\; \langle \prod_{l \in {\cal P}_{min}}\sigma_l \; \rangle_{SU(2)/Z(2)} \right],
\label{mgpred}
\end{equation}
where ${\cal P}_{min}$ is the shortest path between the sites $0$ and $R$.
Next to leading order corrections can be
obtained from (\ref{CFszapp}).

Let us discuss now some
of the issues involved with the above Ising-like effective model as 
well as some possible physical scenarios.

\begin{enumerate}

\item
A link of the above effective model to the vortex condensation theory
presented in the previous section can be established as follows: In
analogy to the $Z(2)$ correlation function let us define the $Z(2)$
vortex free energy. To do this, we rewrite the ratio of partition
functions $q$ defined in (\ref{free2}) as an expectation value of an
appropriate operator which can be expanded in a sum over $SU(2)$
representations. Making use of the decomposition (\ref{5}) - (\ref{o3meas})
and proceeding along the same line as in the case of the correlation
function we are able to express the vortex free energy in terms of
expectation values defined in (\ref{FL}) and (\ref{CFszapp}). Thus, if
it turns out that condensation of vortices is responsible for a
non-zero MG and this MG can be extracted from (\ref{CFsz}) and
(\ref{CFszapp}) respectively, then the $Z(2)$ degrees of freedom
reproduce the exponential decay of the vortex free energy as
well. However, this is only possible in a $SU(2)/Z(2)$ background,
since the $Z(2)$ degrees of freedom alone cannot account for an
exponential fall off.

\item
If $Z(2)$ degrees of freedom indeed play a crucial role in generating
a non-zero MG, this may have some strong impact to $SU(2)/Z(2)$
models. Consider for example the following variant of an $SO(3)$ model
which is an analogue of the lattice gauge model introduced in
\cite{so3}
\begin{eqnarray}
Z^{SO(3)} &=& \sum_{\{s_l\}=\pm 1} \int \prod_x D\mu(U_x) 
\exp [ \, \beta \sum_l s_l \,
\mbox{Tr} (U_x U_{x+n}^\dagger) \, ] =  \nonumber   \\
&=& \int \prod_x D\mu(U_x) \prod_l
\cosh[ \, \beta \, \mbox{Tr} (U_x U_{x+n}^\dagger) \, ].
\label{PFso3}
\end{eqnarray}
At large $\beta$ we obtain
\begin{equation}
Z^{SO(3)} = \int \prod_xD\mu(U_x) \prod_l
\exp [ \, \beta \, | \mbox{Tr} (U_x U_{x+n}^\dagger) | ].
\label{PFso3as}
\end{equation}
Speculations about the coincidence of $SO(3)$ and $SU(2)$ models are
based on the widely accepted belief that the continuum limit has to be
taken at $\beta \to \infty$, where the naive continuum limits of both
models coincide. But since the $SO(3)$ model (\ref{PFso3as}) lacks of
$Z(2)$ degrees of freedom, a possible non-zero MG cannot be explained
with the above described effective model. Thus, either $SU(2)$ and
$SO(3)$ models have different continuum limits, or the Ising-like part
(\ref{PFIsing}) of the $SU(2)$ partition function (\ref{fctrz}) as
well as the correlation function (\ref{CFszapp}) must become trivial
in this limit. The latter scenario means that the MG extracted from
the $Z(2)$ degrees of freedom vanishes above some pseudo-critical value of
$\beta$ while the full $SU(2)$ MG does not. This seems to be very
unlikely, since at least for finite values of $\beta$ the $Z(2)$
correlation function carries the whole MG. This will be shown by
numerical results in the next section.

\item
Suppose for a moment that the $Z(2)$ correlation function indeed 
reproduces the full MG. Comparing formulae (\ref{PFIsing}) 
and (\ref{Isingm}) one can determine an effective Ising 
coupling in the $SU(2)$ model. Two scenarios are possible.
If the effective coupling is less than the critical coupling
of the Ising model for arbitrarily large values of the bare coupling 
$\beta$, the $SU(2)$ model is always in a phase with a 
non-zero MG and the continuum limit may be taken at $\beta\to\infty$
according to the conventional scenario (at the same time
the effective coupling has to approach the critical coupling
of the Ising model, otherwise the very existence of a nontrivial
continuum limit becomes problematic).
It may happen, however, that at some large value of $\beta$ the effective
Ising coupling becomes larger than the critical coupling
of the Ising model. Then above this $\beta$-value the system
is in a massless phase and the conventional scenario is broken
since one has to realize the continuum limit at this finite
$\beta$-value corresponding to the critical effective 
Ising coupling. It is interesting to mention that in this case 
if it is possible to
construct the massless continuum limit by driving the bare 
coupling to its critical value from above, one may expect that
this limit coincides with the continuum limit of the $SO(3)$ model.   

\end{enumerate}

\section{Monte-Carlo study of $Z(2)$ and $SU(2)$ mass gap}

In this section we will show by
results of MC simulations that the long distance
behaviour of the full $SU(2)$ correlation function (\ref{2}) coincides
with the long distance behaviour of the $Z(2)$ correlation function (\ref{z2corr}). This
is done by calculating the full $SU(2)$ correlation length
$\xi_{SU(2)}$ (which is the inverse of the MG $m$) and by comparing
it with the correlation length $\xi_{Z(2)}$ extracted from $Z(2)$ 
degrees of freedom. As will be seen, both quantities - $\xi_{SU(2)}$
and $\xi_{Z(2)}$ - agree within errorbars. This result confirms the
assumption made in the previous section that the $Z(2)$ degrees of
freedom carry the full $SU(2)$ MG. Furthermore, we
consider the PL model and demonstrate that $Z(2)$
excitations reproduce the full $SU(2)$ MG in this model as well.

\subsection{Standard model}

To calculate the full $SU(2)$ correlation length $\xi_{SU(2)}$ we
follow the method presented in \cite{edwards}\footnote[2]{To be precise,
in ref.~\cite{edwards} the correlation length $\xi$ is defined as the
second-moment correlation length and not as the inverse of the mass 
gap $m$ which can be determined by fitting the falloff of the
zero-momentum correlation function to the $\cosh$-behaviour appropriate
for time-periodicity. However, both quantities are expected to show
the same scaling behaviour. Surprisingly, it is found empirically that
the two definitions of $\xi$ do not only scale in the same way but
agree within less than $1 \%$ \cite{edwards,wolff1}. In this article we
employ the definition of $\xi$ presented in \cite{edwards} since it 
is less CPU-time consuming and we feel free to call it the inverse of the
MG.}. We parameterize our field variables $U_x \in SU(2)$ according
to
\begin{equation}
\label{SU2par}
U_x \; = \; u_0(x) + i \vec{u}(x) \vec{\sigma}, \quad \quad \quad u_x
\; = \; (u_0(x),\vec{u}(x)),
\end{equation}
with $\vec{\sigma}$ the Pauli matrices. We then determine the
susceptibility $\chi_{SU(2)}$ according to
\begin{equation}
\label{susc}
\chi_{SU(2)} \; = \; \frac{1}{V} \, \langle \, ( \sum_x u_x )^2 \, \rangle 
\end{equation}
and the analogous quantity at the smallest non-zero momentum
\begin{equation}
\label{suscn0}
F_{SU(2)} \; = \; \frac{1}{V} \, \langle \frac{1}{2} \left[ | \sum_x
e^{2\pi i x_1/L} u_x |^2 + | \sum_x
e^{2\pi i x_2/L} u_x |^2 \right] \rangle. 
\end{equation}
In terms of these two quantities the second-moment correlation
length $\xi_{SU(2)}$ is given by \cite{edwards}
\begin{equation}
\label{corrl}
\xi_{SU(2)} \; = \; \left( \frac{\chi_{SU(2)}/F_{SU(2)}-1}{4\sin^2 
\pi/L} \right)^{1/2}.
\end{equation}
$L$ is the linear size of the lattice and $V=L^2$ the number of sites
in the lattice. In addition, we calculate the internal energy
\begin{equation}
\label{ener}
E_{SU(2)} \; = \; \Gamma_1(\beta) \; = \; \langle
\mbox{Tr}(U_0U_1^\dagger) \rangle
\end{equation}
corresponding to the correlation function of spins separated by
one lattice spacing.
In the case of the $Z(2)$ correlation length $\xi_{Z(2)}$ we consider
the same ensemble of $SU(2)$ degrees of freedom $U_x$ but instead of
calculating the quantities $\chi$ (\ref{susc}) and $F$ (\ref{suscn0}) with
the parameters $u_x$ (\ref{SU2par}) we use the $Z(2)$ degrees of
freedom
\begin{equation}
\label{z2x}
z_x \; = \; \mbox{sign} (\mbox{Tr} U_x) \; = \; \mbox{sign} (u_0(x)).
\end{equation}
$\xi_{Z(2)}$ is then determined by inserting $\chi_{Z(2)}$ and
$F_{Z(2)}$ into (\ref{corrl}) and $E_{Z(2)}$ is given by $\langle
z_0z_1 \rangle$.

In order to simulate the $2D \ SU(2)$ principal chiral model defined
in (\ref{1}) with $N=2$ we use Wolff's cluster algorithm \cite{wolff2}. The model is
considered at different values of the inverse coupling constant
$\beta$. We choose periodic boundary conditions. In all cases we start
with a random configuration and apply at least $10^4$ warm up
sweeps. We generate $5 \cdot 10^5$ configurations and measure the
quantities of interest in every configuration since the cluster
algorithm is known to show small autocorrelation times. To estimate
errorbars we use the jackknife sub-ensemble analysis. In
table~\ref{tablexi} we show the results for $\xi_{SU(2)}$ and
$\xi_{Z(2)}$ for several $\beta$-values and lattice sizes $L=128$ and
$L=256$. The data in the column denoted with $\xi_E$ is taken from
\cite{edwards} and serves as reference results for $\xi_{SU(2)}$. As
is clearly seen from table~\ref{tablexi} the $SU(2)$ correlation
length $\xi_{SU(2)}$ and the $Z(2)$ correlation length $\xi_{Z(2)}$
show perfect agreement within errorbars. This indicates that the
$Z(2)$ degrees of freedom alone carry the full $SU(2)$ MG.
\begin{table}
\begin{center}
\begin{tabular*}{15cm}{c@{\extracolsep\fill} c c c c}
\hline
$L$ & $2\beta$ & $\xi_{SU(2)}$ & $\xi_{Z(2)}$ & $\xi_E$ \\ 
\hline
\\
128 & 2.2 & 14.01 (0.16) & 14.36 (0.24) & 14.02 (0.03) \\  
    & 2.3 & 18.95 (0.13) & 18.66 (0.19) & 18.91 (0.05) \\
    & 2.4 & 25.39 (0.13) & 25.10 (0.18) & 25.15 (0.07) \\
    & 2.5 & 33.14 (0.12) & 32.95 (0.17) & 33.19 (0.07) \\
    & 2.6 & 41.77 (0.12) & 41.32 (0.19) & 41.38 (0.09) \\
    & 2.7 & 49.44 (0.13) & 49.61 (0.19) & 49.60 (0.11) \\
\\
256 & 2.4 & 26.02 (0.37) & 25.23 (0.60) & 25.50 (0.20) \\  
    & 2.5 & 34.51 (0.28) & 34.63 (0.43) & 34.97 (0.16) \\  
    & 2.6 & 46.44 (0.25) & 46.86 (0.38) & 46.66 (0.17) \\  
    & 2.7 & 62.06 (0.25) & 62.28 (0.37) & 61.90 (0.23) \\  
    & 2.8 & 78.20 (0.26) & 78.55 (0.37) & 78.48 (0.27) \\  
\\
\hline
\end{tabular*}
\end{center}
\caption{\label{tablexi}
Estimates for the correlation length $\xi_{SU(2)}$ extracted
from $SU(2)$ degrees of freedom and for $\xi_{Z(2)}$ determined with
the $Z(2)$ degrees of freedom. The data in the right column denoted by $\xi_E$
show results for $\xi_{SU(2)}$ taken from \cite{edwards}.}
\end{table}

For comparison we also computed the internal energy $E$ (\ref{ener})
which corresponds to the correlation function of spins separated by
one lattice spacing. At such a small distance and for large enough
values of $\beta$ the spin system is well ordered. According to a
vortex condensation theory this means that thin vortices are suppressed
and the spin configurations are not $Z(2)$-like at short
distances. Thus, we do {\it not} expect agreement between the full $SU(2)$
internal energy $E_{SU(2)}$ and the internal energy $E_{Z(2)}$
extracted from $Z(2)$ degrees of freedom. Table~\ref{tableen} shows
numerical results which suggest that this is indeed the
case: There is a large discrepancy between $E_{SU(2)}$ and $E_{Z(2)}$
which indicates that the $Z(2)$ degrees of freedom cannot account for
the short distance behaviour of the $SU(2)$ model. 
\begin{table}
\begin{center}
\begin{tabular*}{15cm}{c@{\extracolsep\fill} c c c c}
\hline
$L$ & $2\beta$ & $E_{SU(2)}$ & $E_{Z(2)}$ & $E_E$ \\ 
\hline
\\
128 & 2.2 & 1.24528 (0.00027) & 0.92359 (0.00062) & 1.24509 (0.00001) \\  
    & 2.3 & 1.28420 (0.00018) & 0.95440 (0.00044) & 1.28394 (0.00001) \\
    & 2.4 & 1.31898 (0.00016) & 0.98440 (0.00043) & 1.31886 (0.00001) \\
    & 2.5 & 1.35023 (0.00011) & 1.01017 (0.00042) & 1.35029 (0.00001) \\
    & 2.6 & 1.37882 (0.00009) & 1.03454 (0.00038) & 1.37874 (0.00001) \\
    & 2.7 & 1.40463 (0.00007) & 1.05722 (0.00042) & 1.40460 (0.00001) \\
\\
256 & 2.4 & 1.31541 (0.00358) & 0.98089 (0.00237) &  \\  
    & 2.5 & 1.34857 (0.00171) & 1.00920 (0.00128) & 1.35021 (0.00001) \\  
    & 2.6 & 1.37834 (0.00015) & 1.03407 (0.00034) & 1.37860 (0.00001) \\  
    & 2.7 & 1.40443 (0.00006) & 1.05648 (0.00032) & 1.40443 (0.00001) \\  
    & 2.8 & 1.42805 (0.00005) & 1.07736 (0.00031) & 1.42806 (0.00001) \\  
\\
\hline
\end{tabular*}
\end{center}
\caption{\label{tableen}
Estimates for the internal energy $E_{SU(2)}$ extracted
from $SU(2)$ degrees of freedom and for $E_{Z(2)}$ determined with
the $Z(2)$ degrees of freedom. The data in the right column denoted by $E_E$
show results for $E_{SU(2)}$ taken from \cite{edwards}.}
\end{table}

Finally let us compare our results obtained in the $2D \ SU(2)$
principal chiral model with similar results obtained in $4D$ gauge
theories. In gauge theories, it was shown by numerical simulations
that the asymptotic ST is nearly reproduced by $U(1)$ and
$Z(2)$ degrees of freedom respectively. While $U(1)$ dominance \cite{suzuki}
is in favour of the dual superconductor picture of confinement,
$Z(2)$ dominance \cite{z2max} indicates that a vortex condensation
mechanism is responsible for confinement of static quarks. However, in
both cases the gauge needs to be partially fixed to extract the
relevant degrees of freedom, and this procedure is apparently
ambiguous. It was reported for example that the Abelian
$U(1)$ dominance is convincingly observed only in one particular
gauge, the so-called maximal Abelian gauge. We want to emphasize that
in the $2D \ SU(2)$ principal chiral model considered in this paper,
there is no gauge freedom and thus, there is no need to fix the
gauge. The determination of the $Z(2)$ degrees of freedom (\ref{z2x})
is unambiguous and the same is true for the results presented in
tables~\ref{tablexi} and \ref{tableen}.

\subsection{Positive link model}

The numerical results of the previous section suggest that the $Z(2)$
degrees of freedom alone reproduce the long distance behaviour of the
$SU(2)$ correlation function while this is not the case at short
distances. This observation is in favour of a vortex condensation
theory according to which the $SU(2)$ spin field configurations behave
$Z(2)$-like only at large distances. In this section we investigate
the dependence of this mechanism on the short distance structure of
the $SU(2)$ model. We restrict the trace of the link variables to
positive values and call this model the positive link model
(PLM). According to the conventional scenario the continuum limit of
the $SU(2)$ model has to be taken at $\beta \to \infty$. Thus, the PLM
and the standard $SU(2)$ model have the same continuum
limit. Moreover, a possible mechanism which is responsible for the
existence of a non-zero MG should be the same in both models.

The positive link model is defined as an analogue of the positive
plaquette model in gauge theories \cite{piet}. It restricts the trace
of link variables to positive values and thus suppresses thin vortices
of the order of one lattice spacing. Such thin vortices can be seen as
lattice artifacts. They should not influence the above discussed
vortex condensation mechanism which is based on condensation of thick
vortices having a linear extension of many lattice spacings. In this
sense, the PLM is closer to the continuum than the standard $SU(2)$
model. The action of the PLM can be written as
\begin{equation}
\label{plma}
S_{PLM} \; = \; \beta \sum_{x,n} \, \mbox{Tr}(U_xU_{x,n}^\dagger) \; -
\; \lambda \sum_{x,n} \, [1-\mbox{sign}(\mbox{Tr}(U_xU_{x,n}^\dagger))],
\end{equation}
where for a complete suppression of negative links we have to choose
$\lambda = \infty$. To simulate the partition function $Z_{PLM} = \int
D\mu(U_x) \exp (S_{PLM})$ of the PLM we use a heatbath
algorithm. In the update the change of a $SU(2)$ spin variable is rejected,
if the trace of one or more of the resulting four links is
negative. Fortunately, the rejection rate decreases with increasing
values of $\beta$. Simulations were run on a lattice with size $L=128$
and at four different $\beta$-values. In all cases we started with the
trivial configuration and applied $2 \cdot 10^4$ warm up sweeps. We then
generated $5 \cdot 10^6$ configurations and measured the quantities of
interest in every fifth  configuration. It should be emphasized that
in this paper we are not interested in the question of scaling in the
PLM. Our aim is to find out whether the $Z(2)$ degrees of freedom
carry the full $SU(2)$ MG in the PLM as well. We thus calculated the
correlation lengths $\xi_{SU(2)}$ (\ref{corrl}) and $\xi_{Z(2)}$ as 
introduced in the previous section. The numerical results are shown in
table~\ref{tableplm}. It is clearly seen that at least for large
enough values of $\beta$ the two quantities agree within
errorbars. This shows that in the PLM the $Z(2)$ degrees of freedom reproduce
the long distance behaviour of the $SU(2)$ correlation function as
well. We interpret this result as an indication that also at weak
coupling condensation of thick vortices is the mechanism which leads
to the existence of a non-zero MG. For comparison we computed the
internal energy $E_{SU(2)}$ (\ref{ener}). For reasons given in the
previous section we do not expect that the $SU(2)$ internal energy
$E_{SU(2)}$ and the internal energy $E_{Z(2)}$ extracted from $Z(2)$
degrees of freedom agree. Table~\ref{tableplm} shows that this is
indeed the case.
\begin{figure}
\centerline{\input{hist.tex}}
\caption{\label{hist}Distribution of link traces in the positive link model.}
\end{figure}
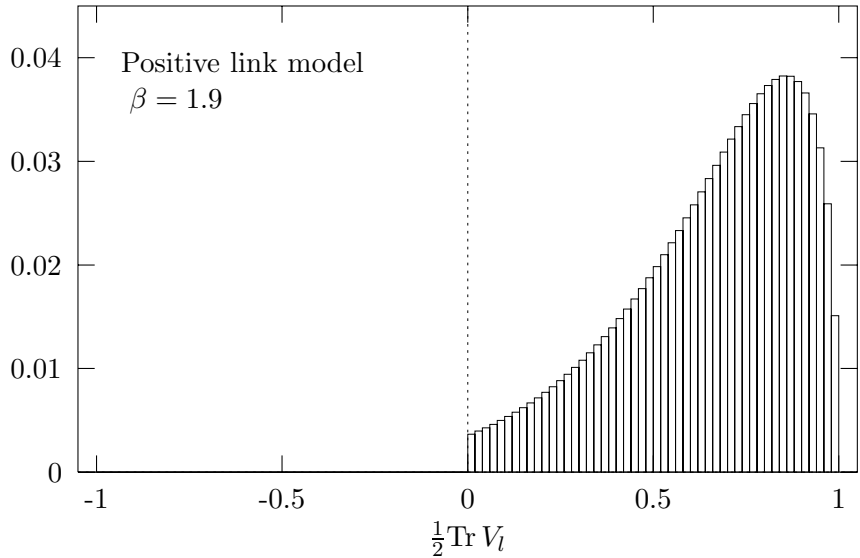

\begin{table}
\begin{center}
\begin{tabular*}{15cm}{c@{\extracolsep\fill} c c c c c}
\hline
$L$ & $2\beta$ & $\xi_{SU(2)}$ & $\xi_{Z(2)}$ & $E_{SU(2)}$ & $E_{Z(2)}$ \\ 
\hline
\\
128 & 1.7 & 20.24 (0.19) & 19.17 (0.40) & 1.27350 (0.00001) & 0.93670 (0.00032) \\  
    & 1.8 & 23.53 (0.21) & 23.92 (0.46) & 1.29278 (0.00001) & 0.95381 (0.00043) \\
    & 1.9 & 26.93 (0.27) & 27.24 (0.55) & 1.31180 (0.00001) & 0.97044 (0.00058) \\ 
    & 2.0 & 31.63 (0.27) & 31.63 (0.94) & 1.33056 (0.00001) & 0.98686 (0.00079) \\ 
\\
\hline
\end{tabular*}
\end{center}
\caption{\label{tableplm}
Estimates for the correlation length $\xi$ and the internal energy $E$
extracted from both $SU(2)$ and $Z(2)$ degrees of freedom in the
positive link model.}
\end{table}

\section{Summary}

In this article we discussed a vortex condensation mechanism to
explain a non-zero MG in the $2D$ $SU(2)$ principal chiral
model. Following the original idea of Mack and Petkova \cite{mack} for
$4D$ non-Abelian gauge theories we formulate a sufficient condition
for the MG to be non-vanishing. This condition is expressed in terms
of the behaviour of the vortex free energy. However, the vortex
condensation mechanism presented here does not specify the definition
of a vortex. The only important issue is that over some characteristic
length scale spin field configurations which are crucial for the
existence of a non-zero MG behave $Z(2)$-like. With this in mind we
separate $Z(2)$ and $SO(3)$ degrees of freedom and construct an
effective model for the $Z(2)$ degrees of freedom in a $SO(3)$
background. Assuming that $Z(2)$ degrees of freedom alone carry the
whole MG we arrive at an effective Ising-like expression for the
$SU(2)$ correlation function. We speculate that due to the non-trivial
$SO(3)$ background the effective Ising-like coupling might be smaller
than the critical Ising coupling for all values of the bare coupling
constant. In order to confirm the above assumption we performed
numerical simulations and compared the full $SU(2)$ MG with the MG
extracted from $Z(2)$ degrees of freedom. We find that $Z(2)$ degrees
of freedom reproduce the full $SU(2)$ MG with perfect accuracy. This
is observed not only in the standard model but also in the positive
link model which due to the complete suppression of links with
negative trace is closer to the continuum limit. We summarize that our
numerical results as well as the effective model for $Z(2)$ degrees of
freedom are in favour of a vortex condensation mechanism.

Finally, we want to stress the following: All arguments presented in
this paper being in favour of a vortex theory did not use
information on the phase structure of the considered model. Thus, they
hold independently of the scenario actually being realized. In
particular, they do not depend on whether the conventional scenario
(with a non-zero MG at arbitrarily large $\beta$ and asymptotic
freedom) or the scenario advocated in ref.~\cite{seiler} (with a phase 
transition to a massless phase at a finite value of $\beta$) is
realized. Moreover, if the conjecture that the $Z(2)$ MG coincides with
the full $SU(2)$ MG is correct, then our effective model might help
to clarify this important question.

\section*{Acknowledgements}

We would like to thank M.~Faber for many stimulating discussions.

\end{document}

%% file: hist.tex
\setlength{\unitlength}{0.1bp}
\special{!
/gnudict 40 dict def
gnudict begin
/Color false def
/Solid false def
/gnulinewidth 2.000 def
/vshift -33 def
/dl {10 mul} def
/hpt 31.5 def
/vpt 31.5 def
/M {moveto} bind def
/L {lineto} bind def
/R {rmoveto} bind def
/V {rlineto} bind def
/vpt2 vpt 2 mul def
/hpt2 hpt 2 mul def
/Lshow { currentpoint stroke M
  0 vshift R show } def
/Rshow { currentpoint stroke M
  dup stringwidth pop neg vshift R show } def
/Cshow { currentpoint stroke M
  dup stringwidth pop -2 div vshift R show } def
/DL { Color {setrgbcolor Solid {pop []} if 0 setdash }
 {pop pop pop Solid {pop []} if 0 setdash} ifelse } def
/BL { stroke gnulinewidth 2 mul setlinewidth } def
/AL { stroke gnulinewidth 2 div setlinewidth } def
/PL { stroke gnulinewidth setlinewidth } def
/LTb { BL [] 0 0 0 DL } def
/LTa { AL [1 dl 2 dl] 0 setdash 0 0 0 setrgbcolor } def
/LT0 { PL [] 0 1 0 DL } def
/LT1 { PL [4 dl 2 dl] 0 0 1 DL } def
/LT2 { PL [2 dl 3 dl] 1 0 0 DL } def
/LT3 { PL [1 dl 1.5 dl] 1 0 1 DL } def
/LT4 { PL [5 dl 2 dl 1 dl 2 dl] 0 1 1 DL } def
/LT5 { PL [4 dl 3 dl 1 dl 3 dl] 1 1 0 DL } def
/LT6 { PL [2 dl 2 dl 2 dl 4 dl] 0 0 0 DL } def
/LT7 { PL [2 dl 2 dl 2 dl 2 dl 2 dl 4 dl] 1 0.3 0 DL } def
/LT8 { PL [2 dl 2 dl 2 dl 2 dl 2 dl 2 dl 2 dl 4 dl] 0.5 0.5 0.5 DL } def
/P { stroke [] 0 setdash
  currentlinewidth 2 div sub M
  0 currentlinewidth V stroke } def
/D { stroke [] 0 setdash 2 copy vpt add M
  hpt neg vpt neg V hpt vpt neg V
  hpt vpt V hpt neg vpt V closepath stroke
  P } def
/A { stroke [] 0 setdash vpt sub M 0 vpt2 V
  currentpoint stroke M
  hpt neg vpt neg R hpt2 0 V stroke
  } def
/B { stroke [] 0 setdash 2 copy exch hpt sub exch vpt add M
  0 vpt2 neg V hpt2 0 V 0 vpt2 V
  hpt2 neg 0 V closepath stroke
  P } def
/C { stroke [] 0 setdash exch hpt sub exch vpt add M
  hpt2 vpt2 neg V currentpoint stroke M
  hpt2 neg 0 R hpt2 vpt2 V stroke } def
/T { stroke [] 0 setdash 2 copy vpt 1.12 mul add M
  hpt neg vpt -1.62 mul V
  hpt 2 mul 0 V
  hpt neg vpt 1.62 mul V closepath stroke
  P  } def
/S { 2 copy A C} def
end
}
\begin{picture}(3600,2160)(0,0)
\special{"
gnudict begin
gsave
50 50 translate
0.100 0.100 scale
0 setgray
/Helvetica findfont 100 scalefont setfont
newpath
-500.000000 -500.000000 translate
LTa
480 251 M
2937 0 V
-1468 0 R
0 1758 V
LTb
480 251 M
63 0 V
2874 0 R
-63 0 V
480 642 M
63 0 V
2874 0 R
-63 0 V
480 1032 M
63 0 V
2874 0 R
-63 0 V
480 1423 M
63 0 V
2874 0 R
-63 0 V
480 1814 M
63 0 V
2874 0 R
-63 0 V
550 251 M
0 63 V
0 1695 R
0 -63 V
1249 251 M
0 63 V
0 1695 R
0 -63 V
1949 251 M
0 63 V
0 1695 R
0 -63 V
2648 251 M
0 63 V
0 1695 R
0 -63 V
3347 251 M
0 63 V
0 1695 R
0 -63 V
480 251 M
2937 0 V
0 1758 V
-2937 0 V
480 251 L
LT0
550 251 M
28 0 V
-28 0 V
28 0 R
28 0 V
-28 0 V
28 0 R
28 0 V
-28 0 V
28 0 R
28 0 V
-28 0 V
28 0 R
28 0 V
-28 0 V
28 0 R
28 0 V
-28 0 V
28 0 R
28 0 V
-28 0 V
28 0 R
28 0 V
-28 0 V
28 0 R
28 0 V
-28 0 V
28 0 R
28 0 V
-28 0 V
28 0 R
28 0 V
-28 0 V
28 0 R
28 0 V
-28 0 V
28 0 R
28 0 V
-28 0 V
28 0 R
28 0 V
-28 0 V
28 0 R
28 0 V
-28 0 V
28 0 R
27 0 V
-27 0 V
27 0 R
28 0 V
-28 0 V
28 0 R
28 0 V
-28 0 V
28 0 R
28 0 V
-28 0 V
28 0 R
28 0 V
-28 0 V
28 0 R
28 0 V
-28 0 V
28 0 R
28 0 V
-28 0 V
28 0 R
28 0 V
-28 0 V
28 0 R
28 0 V
-28 0 V
28 0 R
28 0 V
-28 0 V
28 0 R
28 0 V
-28 0 V
28 0 R
28 0 V
-28 0 V
28 0 R
28 0 V
-28 0 V
28 0 R
28 0 V
-28 0 V
28 0 R
28 0 V
-28 0 V
28 0 R
28 0 V
-28 0 V
28 0 R
28 0 V
-28 0 V
28 0 R
28 0 V
-28 0 V
28 0 R
28 0 V
-28 0 V
28 0 R
28 0 V
-28 0 V
28 0 R
28 0 V
-28 0 V
28 0 R
28 0 V
-28 0 V
28 0 R
28 0 V
-28 0 V
28 0 R
28 0 V
-28 0 V
28 0 R
28 0 V
-28 0 V
28 0 R
28 0 V
-28 0 V
28 0 R
28 0 V
-28 0 V
28 0 R
28 0 V
-28 0 V
28 0 R
28 0 V
-28 0 V
28 0 R
28 0 V
-28 0 V
28 0 R
28 0 V
-28 0 V
28 0 R
28 0 V
-28 0 V
28 0 R
28 0 V
-28 0 V
28 0 R
28 0 V
-28 0 V
28 0 R
28 0 V
-28 0 V
28 0 R
0 143 V
27 0 V
0 -143 V
-27 0 V
27 0 R
0 155 V
28 0 V
0 -155 V
-28 0 V
28 0 R
0 167 V
28 0 V
0 -167 V
-28 0 V
28 0 R
0 180 V
28 0 V
0 -180 V
-28 0 V
28 0 R
0 195 V
28 0 V
0 -195 V
-28 0 V
28 0 R
0 210 V
28 0 V
0 -210 V
-28 0 V
28 0 R
0 226 V
28 0 V
0 -226 V
-28 0 V
28 0 R
0 243 V
28 0 V
0 -243 V
-28 0 V
28 0 R
0 261 V
28 0 V
0 -261 V
-28 0 V
28 0 R
0 280 V
28 0 V
0 -280 V
-28 0 V
28 0 R
0 301 V
28 0 V
0 -301 V
-28 0 V
28 0 R
0 322 V
28 0 V
0 -322 V
-28 0 V
28 0 R
0 345 V
28 0 V
0 -345 V
-28 0 V
28 0 R
0 369 V
28 0 V
0 -369 V
-28 0 V
28 0 R
0 395 V
28 0 V
0 -395 V
-28 0 V
28 0 R
0 422 V
28 0 V
0 -422 V
-28 0 V
28 0 R
0 450 V
28 0 V
0 -450 V
-28 0 V
28 0 R
0 480 V
28 0 V
0 -480 V
-28 0 V
28 0 R
0 511 V
28 0 V
0 -511 V
-28 0 V
28 0 R
0 544 V
28 0 V
0 -544 V
-28 0 V
28 0 R
0 579 V
28 0 V
0 -579 V
-28 0 V
28 0 R
0 615 V
28 0 V
0 -615 V
-28 0 V
28 0 R
0 653 V
28 0 V
0 -653 V
-28 0 V
28 0 R
0 692 V
28 0 V
0 -692 V
-28 0 V
28 0 R
0 733 V
28 0 V
0 -733 V
-28 0 V
28 0 R
0 775 V
28 0 V
0 -775 V
-28 0 V
28 0 R
0 820 V
28 0 V
0 -820 V
-28 0 V
28 0 R
0 865 V
28 0 V
0 -865 V
-28 0 V
28 0 R
0 911 V
28 0 V
0 -911 V
-28 0 V
28 0 R
0 959 V
28 0 V
0 -959 V
-28 0 V
28 0 R
0 1008 V
28 0 V
0 -1008 V
-28 0 V
28 0 R
0 1057 V
28 0 V
0 -1057 V
-28 0 V
28 0 R
0 1107 V
28 0 V
0 -1107 V
-28 0 V
28 0 R
0 1157 V
28 0 V
0 -1157 V
-28 0 V
28 0 R
0 1207 V
28 0 V
0 -1207 V
-28 0 V
28 0 R
0 1256 V
27 0 V
0 -1256 V
-27 0 V
27 0 R
0 1303 V
28 0 V
0 -1303 V
-28 0 V
28 0 R
0 1348 V
28 0 V
0 -1348 V
-28 0 V
28 0 R
0 1390 V
28 0 V
0 -1390 V
-28 0 V
28 0 R
0 1427 V
28 0 V
0 -1427 V
-28 0 V
28 0 R
0 1458 V
28 0 V
0 -1458 V
-28 0 V
28 0 R
0 1481 V
28 0 V
0 -1481 V
-28 0 V
28 0 R
0 1494 V
28 0 V
0 -1494 V
-28 0 V
28 0 R
0 1493 V
28 0 V
0 -1493 V
-28 0 V
28 0 R
0 1473 V
28 0 V
0 -1473 V
-28 0 V
28 0 R
0 1430 V
28 0 V
0 -1430 V
-28 0 V
28 0 R
0 1351 V
28 0 V
0 -1351 V
-28 0 V
28 0 R
0 1223 V
28 0 V
0 -1223 V
-28 0 V
28 0 R
0 1012 V
28 0 V
0 -1012 V
-28 0 V
28 0 R
0 590 V
28 0 V
0 -590 V
-28 0 V
currentpoint stroke M
stroke
grestore
end
showpage
}
\put(1100,1800){\makebox(0,0){Positive link model}}
\put(850,1650){\makebox(0,0){$\beta=1.9$}}
\put(1948,0){\makebox(0,0){$\frac{1}{2}\mbox{Tr} \, V_l$}}
\put(3347,151){\makebox(0,0){1}}
\put(2648,151){\makebox(0,0){0.5}}
\put(1949,151){\makebox(0,0){0}}
\put(1249,151){\makebox(0,0){-0.5}}
\put(550,151){\makebox(0,0){-1}}
\put(420,1814){\makebox(0,0)[r]{0.04}}
\put(420,1423){\makebox(0,0)[r]{0.03}}
\put(420,1032){\makebox(0,0)[r]{0.02}}
\put(420,642){\makebox(0,0)[r]{0.01}}
\put(420,251){\makebox(0,0)[r]{0}}
\end{picture}